\documentclass[twocolumn,showpacs,amsmath,amssymb,prl]{revtex4}%superscriptaddress
\usepackage{graphicx}

\begin{document}

% Use the \preprint command to place your local institutional report
% number in the upper righthand corner of the title page in preprint mode.
% Multiple \preprint commands are allowed.
% Use the 'preprintnumbers' class option to override journal defaults
% to display numbers if necessary
%\preprint{}
%Title of paper
\title{Tuning phase transitions of FeSe thin flakes by field effect transistor with solid ion conductor as gate dielectric}

\author{B. Lei$^1$, N. Z. Wang$^1$, C. Shang$^1$, F. B. Meng$^1$, L. K. Ma$^1$, X. G. Luo$^{1,4}$, T. Wu$^{1,4}$, Z. Sun$^{2,4}$, Y. Wang$^5$, Z. Jiang$^5$, B. H. Mao$^6$, Z. Liu$^{6,7}$, Y. J. Yu$^8$, Y. B. Zhang$^{4,8}$}
\email{zhyb@fudan.edu.cn}
\author{X. H. Chen$^{1,3,4}$}
\email{chenxh@ustc.edu.cn}

\affiliation{ $^1$Hefei National Laboratory for Physical Sciences at
Microscale and Department of Physics, and CAS Key Laboratory
of Strongly-coupled Quantum Matter Physics, University of Science and
Technology of China, Hefei, Anhui 230026, China\\
$^2$National Synchrotron Radiation Laboratory, University of Science
and Technology of China, Hefei, Anhui 230026, China\\
$^3$High Magnetic Field Laboratory, Chinese Academy of Sciences,
Hefei, Anhui 230031, China\\
$^4$Collaborative Innovation Center of Advanced Microstructures,
Nanjing University, Nanjing 210093, China\\
$^5$Shanghai Institute of Applied Physics, Chinese Academy of
Sciences, Shanghai 201204, China\\
$^6$State Key Laboratory of Functional Materials for Informatics,
Shanghai Institute of Microsystem \\ and Information Technology,
Chinese Academy of Sciences, Shanghai 200050, China\\
$^7$School of Physical Science and Technology, Shanghai Tech
University, Shanghai 200031, China\\
$^8$State Key Laboratory of Surface Physics and Department of
Physics, Fudan University, Shanghai 200433, China}

\date{\today}

\begin{abstract}
We develop a novel field effect transistor (FET) device using solid ion conductor (SIC) as a
gate dielectric, and we can tune the carrier density of FeSe by
driving lithium ions in and out of the FeSe thin flakes, and
consequently control the material properties and its phase
transitions. A dome-shaped superconducting phase diagram was mapped
out with increasing Li content, with $T_c$ $\sim$ 46.6 K for the
optimal doping, and an insulating phase was reached at the extremely
overdoped regime. Our study suggests that, using solid ion conductor
as a gate dielectric, the SIC-FET device can achieve much higher
carrier doping in the bulk, and suit many surface sensitive
experimental probes, and can stabilize novel structural phases that
are inaccessible in ordinary conditions.

\end{abstract}

% insert suggested PACS numbers in braces on next line


%\maketitle must follow title, authors, abstract, \pacs, and \keywords
\maketitle

Chemical doping is a conventional way to introduce charge carriers
into solids by replacing one of the constituent elements with
another element of a different valence state. For instance, high
temperature superconductivity is realized by suppressing the
antiferromagnetism or spin density wave with chemical doping of
~10\% or ~10$^{21}$ dopant atoms per cm$^3$ in copper oxides and
iron-based superconductors \cite{18,19,20}. However, the chemical
doping is incapable in many cases, because the element replacement
and the variation of carrier density cannot practically cover a
large regime and leave many phases unexplored. As a complementary
method, the application of field effect transistors (FET) in
two-dimensional systems is an effective way to control electronic
properties via reversible changes of charge carrier density
\cite{1,2,3,4,5,6,7,8,9,10,11,12,13,14,15}. Such an electrostatic
doping is desirable to study novel phases that cannot be achieved by
material synthetic methods \cite{6,7,8,10,12,13,4,15}. For instance,
we have utilized tunable ion intercalation with an ionic liquid to
alter charge-ordered states in 1T-TaS$_2$ and induce phase
transitions in thin flakes with reduced dimensionality \cite{12}.
The FET devices have been widely applied in the exploration of new
superconductors \cite{7,8}, the preparation for new devices
\cite{21,22} as well as many applications in semiconductor industry
\cite{4}.

So far, only two types of field effect transistor (FET) devices,
metal-insulator-semiconductor (MIS) FET (Fig. 1(a)) and electric
double layer (EDL) FET (Fig. 1(b)), can be widely used to
continuously tune carrier density \cite{16,17}. However, there are
inherent drawbacks for the investigation of novel quantum phases in
solids. In conventional MIS-FET devices, the electrostatic doping to
a system is realized by the accumulation of mobile carriers at the
surface of an insulator with a gate voltage applied
\cite{3,4,8,16,17}. A limited sheet carrier density n$_{2D}$ of
$\sim$1 $\times$ 10$^{13}$ cm$^{-2}$ can be obtained on the surface
of an insulator before the gate dielectric breaks down due to the
large electric field \cite{2,6,8}. They cannot provide sufficient
carriers to induce superconductivity \cite{17}. The EDL-FET with a
liquid electrolyte as a gate dielectric can achieve a higher
two-dimensional carrier density n$_{2D}$ of $\sim$8 $\times$
10$^{14}$ cm$^{-2}$ \cite{23}. However, the overlay of liquid
electrolyte makes it difficult to suit the modern electronic
technology and prevents heavily-doped electronic states from being
characterized by many physical measurements. In addition, many
materials of interest have electrochemical reaction with liquid
electrolyte \cite{24,25}. Moreover, both of FETs can only tune the
accumulation of carrier on the surface of materials. In this letter,
using solid ion conductor as a gate dielectric, we introduce a new
type of FET, SIC-FET (Fig. 1(c)). This type of FET devices solves
the shortcoming of the carrier control methods mentioned above and
can pave the way for the investigation of new electronic states in
solids.

\begin{figure}[h]
    \centering
    \includegraphics[width=0.42\textwidth]{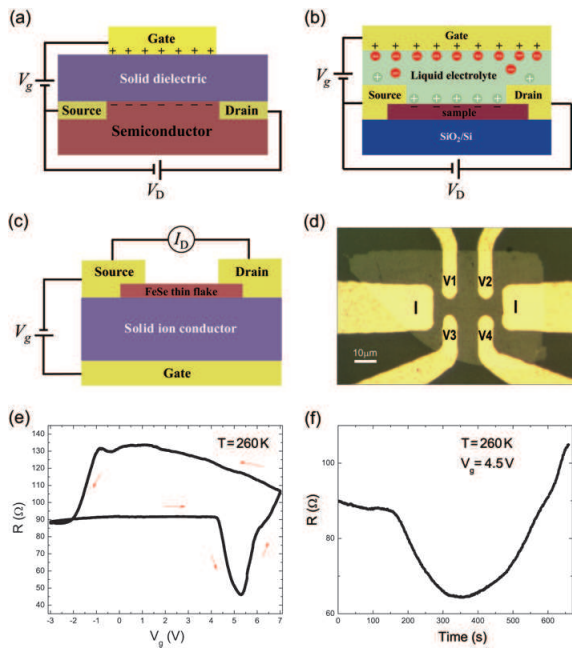}
    \caption{(color online). Resistance controlled by gate voltage with SIC-FET device. (a) A schematic illustration of the MIS-FET with solid oxide such as SiO$_2$ or SrTiO$_3$ as the gate dielectric; (b) A schematic structure of EDL-FET with liquid electrolyte as the gate dielectric; (c) A schematic diagram of the SIC-FET device with the solid ion conductor as the gate dielectric; (d) The optical image of a FeSe thin flake with a standard Hall bar configuration with the current and voltage terminals labeled; (e) Gate voltage dependence of the resistance for a FeSe thin flake with typical thickness of about 18 nm in SIC-FET device. The continuously swept gate voltage with a scan rate of 2 mVs$^{-1}$ was applied at 260 K; (f) The evolution of the resistance with time in SIC-FET with V$_g$ = 4.5 V where the resistance of the sample started to drop. }
    \label{Fig. 1}
\end{figure}

To demonstrate the capability of the SIC-FET device, we choose FeSe
as a model material. FeSe has the simplest structure in iron-based
high-temperature superconductors, with conducting FeSe layers
stacked along c-axis \cite{26}. FeSe and its derived superconductors
are currently the focus of research in the field of iron-based
superconductors \cite{27,28,29,30,31}. In particular, the monolayer
FeSe thin film on SrTiO$_3$ substrate has generated wide research
interest because of its unexpected high- $T_c$ superconductivity
with $T_c$ as high as 65 K, in sharp contrast to the bulk $T_c$ $<$
10 K \cite{27,31,32,33,34,35,36}. Numerous studies have shown that
those FeSe-derived materials with $T_c$ $>$ 40 K are heavily
electron-doped systems, which share very similar electronic
structures \cite{37,38}. The origin of superconductivity in these
high-$T_c$ materials remains to be an intriguing topic.

Recently, we have systematically tuned the superconductivity of FeSe
thin flakes by electron doping with a FET device \cite{39}. A onset
high-$T_c$ of 48 K was achieved in FeSe thin flakes with initial
$T_c$ less than 10 K. Intriguingly, a Lifshitz transition occurs at
a certain carrier concentration, leading to a sudden change of $T_c$
from less than 10 K to more than 30 K \cite{39}. However, the
superconducting regime in the carrier-doping phase diagram is
incomplete due to the sample damage caused by electrochemical
reactions between samples and ionic liquid at high gate voltage
($V_g$ $>$ 6 V). To investigate the whole superconducting regime, a
more effective method is required to introduce higher carrier
density into FeSe.

Using the newly developed SIC-FET devices to increase electron
doping in FeSe thin flakes by driving Li$^+$ ions into samples, we
have found structural transitions from FeSe (11) phase to a
low-doping Li$_y$Fe$_2$Se$_2$ (122 phase-I), then to a high-doping
122 phase-II. A dome-shaped superconducting phase diagram was mapped
out with increasing Li content, and $T_c$ is enhanced from 8 K in
FeSe to 46.6 K for the optimal doping, then decreases in the
overdoped regime and eventually an insulating phase emerges. The
SIC-FET device proves to be able to introduce much higher carrier
doping in bulk materials and stabilize novel structural phases. The
application of such a novel FET device can provide exciting
opportunities for exploring new quantum phases and new materials.

We prepared FET devices with the solid state lithium ion conductive
glass ceramics as the gate dielectrics. The exfoliated FeSe thin
flakes with a typical thickness of $\sim$18 nm were used to
fabricate the transport channel. Li$^+$ in the lithium ion conductor
can move under the applied electric field. For positive gate
voltage, Li$^+$ accumulates on the surface of samples, and then
enters into samples to tune the carrier concentration. Figure 1(c)
depicts a schematic illustration of the SIC-FET device. Detailed
device preparation procedures are described in the Supplemental
Material. Fig. 1(d) shows the optical image of a FeSe thin flake
with a standard Hall bar configuration and with current and voltage
terminals labeled. A continuously swept positive gate voltage with a
scan rate of 2 mVs$^{-1}$ was applied at 260 K, which is the optimal
temperature for applying gate voltages. A typical $R$-$V_g$ curve is
shown in Fig. 1(e). The resistance of the FeSe thin flake remains
almost unchanged with $V_g$ $<$ 4.5 V, and starts to drop quickly
for $V_g$ $ > $ 4.5 V, and reaches a minimum at 5.3 V, then
increases rapidly. When the gate voltage is swept back to -2 V, the
resistance can recover to the initial value. This behavior indicates
that the tuning process is reversible. In fact, when the resistance
of the sample starts to drop, the process of Li$^+$ intercalating
into the FeSe thin flake sample will continue even if the gate
voltage stays at a certain level. As shown in Fig. 1(f), a
continuous change of the resistance of a sample in the relaxation
process at a fixed gate voltage and temperature is quite similar to
that with continuous sweeping of the gate voltage.

In FeSe-derived superconductors, numerous studies suggest that
carrier concentration doped into FeSe layers is the primary factor
that controls the superconducting transition temperature. When
Li$^+$ ions in lithium ion conductor is driven to a FeSe think flake
with gating electric field, we can effectively introduce electron
carriers into FeSe layers. After intercalating Li$^+$ into FeSe thin
flakes at 260 K, we quickly cooled down the FET devices to 210 K,
below which the mobility of Li$^+$ ions in thin flake samples is
completely suppressed. The gate voltage of 4.5 V, at which the
resistance of sample starts to drop, was kept fixed during the whole
measurements. In order to achieve a fine modulation of Li$^+$ ions
intercalation, the FET device was relaxed at 260 K for a certain
period of time before cooling down in each round. This procedure can
lead to a series of doping levels with a controllable fashion.

\begin{figure}[h]
    \centering
    \includegraphics[width=0.45\textwidth]{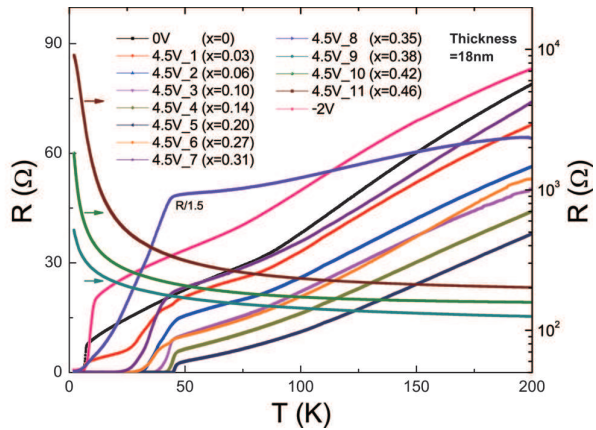}
    \caption{(color online). Temperature dependence of resistance for a FeSe thin flake without gating and with gate voltage V$_g$=4.5 V. The thickness of the thin flake determined by atomic force microscopy is 18 nm. The \textit{x} is Li:Fe ratio whose determination is discussed in the Supplemental Material. }
    \label{Fig. 2}
\end{figure}

Using the four-terminal configuration, we measured a FeSe thin flake
with various amounts of intercalated Li$^+$ ions below 200 K and
show them in Fig. 2. At $V_g$ = 0 V, the sample is superconducting
with the onset critical temperature $T_c^{onset}$= 7.3 K and reaches
zero-resistance at 5 K. Here, $T_c^{onset}$ is defined as the
intersection between the linear extrapolation of the normal state
and the superconducting transition. With increasing the amount of
intercalated Li$^+$ ions, the electronic carrier density of the
sample increases (as shown in Fig. S2(b) of Supplemental Material),
and the $T_c$ as a function of Li$^+$ concentration shows a
dome-like behavior. For the fifth round of cooling down at $V_g$ =
4.5 V, the optimal superconductivity of the FeSe thin flake has been
obtained with the Li/Fe ratio of 0.2. With further increasing the
amount of intercalated Li$^+$ ions, the $T_c$ starts to decrease,
and eventually the FeSe thin flake shows an insulating behavior.
When the gate voltage is swept back to -2 V, the resistance comes
back to the initial value and the $T_c$ also turns back to about 8
K, being almost the same as the initial $T_c$ before gating. This
behavior further indicates that the carrier tuning procedure with
the SIC-FET device is reversible, though the evident higher
resistance indicates that the intercalation of Li$^+$ ions into the
system may damage the sample in some way. We note here that both the
optimal $T_c^{onset}$=46.6 K and the transition from the
superconductivity to insulating behavior were repeatable in every
measured device. The sample with optimal doping exhibits
$T_c^{onset}$=46.6 K and zero-resistance temperature $T_c^{zero}$ of
44.8 K, which is slightly higher than that of Li/ammonia
intercalated FeSe synthesized by the ammonothermal method \cite{29}.
The width of the superconducting transition is less than 2 K, much
sharper than that of $\sim$ 13 K observed by tuning carrier
concentration with EDL-FET device for FeSe thin flakes and thin
films \cite{39,40} (see Fig. S3). This contrast suggests that the
carrier concentration distribution in the SIC-FET devices is more
homogeneous than that in the EDL-FET devices.

Owing to the small ionic radius of Li$^+$ ions, they can be easily
intercalated into layered materials and alter local crystal
structures of materials. Figure 3 shows the \textit{in-situ} X-ray
diffraction (XRD) patterns for a FeSe thin flake with thickness of
40 nm before and after the intercalation of Li$^+$ ions. Before the
intercalation of Li$^+$, the diffraction peak of (001) appears at
2$\theta$=16.09$^{\circ}$. With increasing the amount of
intercalated Li$^+$ ions, the (001) peak intensity shows a drastic
decrease with no noticeable variation in position. The phase with
extremely low amount of Li$^+$ ions possesses the same structure as
that of FeSe (11 phase), in which Li$^+$ ions diffuse randomly in
the FeSe crystal. With the increase of Li$^+$ ions, two additional
diffraction peaks appear at lower angles 13.76$^{\circ}$ and
13.29$^{\circ}$, indicating new phases derived from the structure of
FeSe. These new peaks do not belong to the same phase because only
the 13.29$^{\circ}$ peak shows up in the insulating phase (see Fig.
3b). The data suggests that the intercalation of Li$^+$ ions into
FeSe thin flakes results in two new phases.

\begin{figure}[h]
    \centering
    \includegraphics[width=0.45\textwidth]{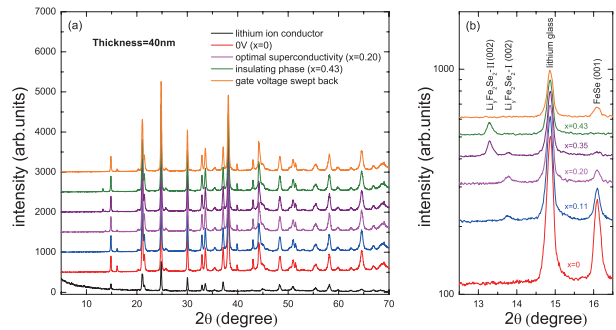}
    \caption{(color online). \textit{In-situ} X-ray diffraction (XRD) patterns for the FeSe thin flake in SIC-FET device before and after the intercalation of Li$^+$ ions. (a) The typical XRD patterns of the FeSe thin flakes with thickness of 40 nm for different charge stages; (b) The magnified view of the angle range from 12.5$^{\circ}$ to 16.5$^{\circ}$.}
    \label{Fig. 3}
\end{figure}

To identify the phases with stronger signals, we carried out XRD
measurements on a thicker flake of 150 nm. As shown in Fig. S5 in
the Supplemental Material, two sets of (00\textit{l}) diffraction
peaks have been identified, suggesting that the separations between
adjacent Fe layers (\textit{d}) are 6.4128 \AA and 6.6570 \AA,
respectively. In the Na$_{0.65} $Fe$ _{1.93} $Se$ _{2} $ and K$ _{x}
$Fe$ _{2-y} $Se$ _{2} $ of ThCr$ _{2} $Si$ _{2} $-type structure,
the separations between neighboring FeSe layers are 6.8339 \AA and
7.0180 \AA, respectively \cite{27,41}. Considering that the ionic
radius of Li$^+$ is smaller than that of Na$^+$ and K$^+$, we
attribute the two new phases to the ThCr$ _{2} $Si$ _{2} $-type
structure. We label the phase with a smaller lattice parameter
\textit{c} (12.8256 \AA) as Li$ _{y} $Fe$ _{2} $Se$ _{2} $-I (122-I)
phase and the one with larger \textbf{c} (13.314 \AA) as Li$ _{y}
$Fe$ _{2} $Se$ _{2} $-II (122-II) phase. The different layer
separation d in the two phases can be attributed to the different
Li$^+$ contents. Generally speaking, the small cation may have more
diverse arrangements and coordination environments. Similar multiple
phases have been reported in Na-intercalated FeSe superconductors
\cite{41}. When the thin flake is tuned to an insulator, the 11
phase and 122-I phase completely disappear and only 122-II phase
exists as shown in Fig. 3. When the gate voltage is swept back to -2
V, the 122-I and -II phases completely disappear, and the FeSe phase
recovers. This result indicates that the newly formed 122 structural
phases can only exist under the electric field, and also implies
that the process of intercalating Li$^+$ into the FeSe controlled by
electric field is reversible. We note that the two new
Li-intercalated FeSe phases have not been reported before, which are
induced and stabilized by the SIC-FET device. Since the
superconductivity in bulk FeSe-derived materials is primarily
determined by carrier concentration doped into FeSe layers, the
emerging new 122 structures may not play a primary role in the novel
electronic properties of FeSe layers. However, the superconductivity
and other properties, and especially the insulating behavior of
these new 122 phases is very intriguing for further investigation.

To acquire more details about the electron doping and structural
modification, \textit{in-situ} X-ray photoelectron spectroscopy
(XPS) and X-ray absorption near edge spectra (XANES) were performed
on the SIC-FET devices to study FeSe thin flakes with different
lithium contents. The XPS result for Fe \textit{2p} is shown in Fig.
S5. The spectrum taken on FeSe single crystals agrees well with that
of the as-grown FeSe film \cite{42}. It consists of two features at
706.6 eV and 707.8 eV. We use the center of the weight of the Fe
\textit{2p} peaks to qualitatively describe the change of Fe
chemical valence. The center point decreases from 707.1 eV (no
gating), 707.0 eV (optimal superconductivity) and eventually to
706.6 eV (insulating phase). Clearly, the Fe valence state decreases
with the increase of Li$^+$ ions intercalation. Due to the lack of
reference data and surface sensitive nature of XPS, it is difficult
for us to determine the Fe valance accurately at this moment. In
addition, the XANES Fe \textit{K}-edge spectra are shown in Fig. S6.
This main absorption feature is due to the transition from Fe
\textit{1s} to the \textit{4p} state. The absorption edge of the
insulating phase FeSe thin flake shifts 2.5 eV towards lower energy
relative to that of pure FeSe, corresponding to a change of the Fe
valence state from +2 in the pure FeSe to 1.23$\pm$0.15 in the
insulating phase. Furthermore, the slight parallel shift of
absorption edge at the optimal superconducting state gives
additional evidence for the random distribution of Li over Fe since
a new local ordering may give rise to the deviations in the
absorption spectra. On the other hand, the distinct shift of the
absorption edge in the insulating phase from that of pure FeSe
sample indicates a new ordering or structural change due to Li
intercalation. Such a change is also responsible for the variation
of the pronounced pre-edge feature at 7111 eV, which arises from the
quadruple transition from Fe \textit{1s} to Fe \textit{3d}. These
findings support the formation of a new 122 structure phase in the
insulating state. The decrease of Fe valence state observed in both
of XPS and XANES suggests that Li$^+$ intercalation controlled by
electric field is an effective method to introduce electron carriers
into FeSe layers. The XANES results indicate that the Li is randomly
distributed in 122-I phase, while is ordered in 122-II phase.
Further investigation is required to determine Li positions and to
reveal their microstructures in the two phases.

\begin{figure}[h]
    \centering
    \includegraphics[width=0.45\textwidth]{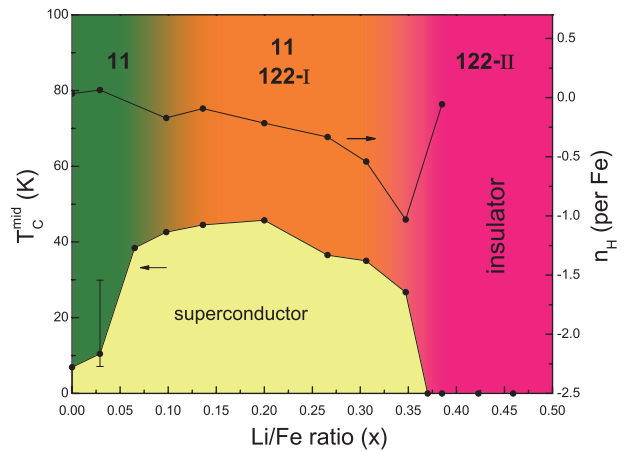}
    \caption{(color online). The phase diagram of Li-intercalated FeSe thin flake as a function of Li/Fe ratio. The $T_c$ shows a dome-like behavior with increasing Li/Fe ratio. A phase transition from superconductor to insulator happens around Li/Fe ratio of 0.37. A series of structural phase transitions from FeSe 11 phase to Li$ _{y} $Fe$ _{2} $Se$ _{2} $-I phase, then to Li$ _{y} $Fe$ _{2} $Se$ _{2} $-II phase takes place. The Hall number n$_H$ is plotted to show the variation of carrier density.  }
    \label{Fig. 4}
\end{figure}

The phase diagram of superconductivity and phase variations of
Li$^+$ intercalated FeSe thin flakes are plotted as a function of
Li/Fe ratio in Figure 4. The method of determining the amount of
intercalated Li$^+$ content is described in the Supplemental
Material. Based on the experimental results, the SIC-FET not only
tunes the carrier concentration of the samples, but also modulates
the crystal structure of materials. With increasing the amount of
intercalated Li$^+$, the $T_c$ shows a dome-like behavior. The
optimal superconductivity with mid-point critical temperature
$T_c^{mid}$=45.8 K has been achieved when the Li/Fe ratio reaches
0.2. As the ratio is up to about 0.37, the superconductivity
disappears completely, and then the sample becomes an insulator. A
similar evolution from superconductor to insulator has been also
observed in (Li,Fe)OHFeSe thin flake by electric field controlled
lithium doping \cite{43}. Most intriguingly, an abrupt jump of the
$T_c$ from $\sim$10 K to $\sim$40 K is observed when the Li/Fe ratio
is around 0.07. Meanwhile, the Hall number n$_H$ shows a
corresponding sudden sign reversal. The evolution of Hall signal
from hole-type to electron-type upon the increase of electron
concentration induced by Li$^+$ intercalation can serve as an
indicative of Lifshitz transition \cite{44}. Similar transition has
been observed in FeSe thin flakes with EDL-FET device \cite{39}.
However, for a comparison, the effective carrier concentration
introduced by Li$^+$ intercalation can further increase and go
beyond the upper limit of EDL-FET devices. As shown in Fig. 4, the
carrier concentration suddenly decreases when the sample is tuned to
an insulator, which could be attributed to some charge localization
effect. In the whole phase diagram, the system experiences a series
of phase transitions. In the beginning of Li$^+$ intercalation into
FeSe thin flakes, the system keeps the same structure as FeSe, and a
sudden enhancement of $T_c$ occurs. With further increasing Li$^+$
intercalation, the 122-I phase emerges and coexists with the FeSe
phase, then 122-II phase shows up at higher doping level. When the
system is tuned to an insulator, FeSe phase and 122-I phase
completely disappear, and only 122-II phase remains. We note that
the coexistence of the 11, 122-I and -II phases is not intrinsic due
to the inhomogeneity of the Li$^+$ distribution. Such inhomogeneous
distribution of Li$^+$ in thin flakes strongly depends on the
thickness of thin flakes, which was confirmed by the XRD patterns
shown in Fig. 3 and Fig.S5 for thin flakes with different thickness
of 40 nm and 150 nm, respectively. It should be pointed out that the
data shown in Fig. 2 were obtained on the FeSe thin flakes of 18 nm,
which should be much more homogeneous than that observed in Fig. 3.
It is possible that no coexistence of the 11, 122-I and -II phases
takes place in the devices with the flakes of 18 nm. We have tried
to perform XRD measurements on the device with the 18 nm thin flakes
by XRD, but the sample is too thin to detect signal. We thus
selected a 40 nm FeSe thin flake for XRD characterization, which
could inevitably involve inhomogeneity.

In summary, using solid lithium ion conductor as the gate
dielectric, we developed a novel FET device (SIC-FET), which can
effectively introduce electron carriers into layered materials.
Compared to the MIS-FET and the EDL-FET devices, the SIC-FET device
shows stronger capability of tuning carrier concentration, so that
it can help to map out the phase diagram of FeSe over a wide range
for the first time. Owing to the small ionic radius of Li$^+$, the
SIC-FET device can tune the carrier concentration of entire
materials with thickness of ten nanometers. Besides, the SIC-FET
device can be used to search for new materials or novel
superconductors with metastable structures. In particular, the
configuration of SIC-FET devices is suitable for many other
surface-sensitive experimental examinations. In our studies, two
novel Li-intercalated FeSe phases with ThCr$ _{2} $Si$ _{2} $-type
structure, which cannot be synthesized by conventional methods, were
obtained under the electric field. The investigation on the
insulating Li$ _{y} $Fe$ _{2} $Se$ _{2}$-II phase could be helpful
for the understanding of the superconducting mechanism for
FeSe-derived superconductors. Our findings demonstrate the potential
of the SIC-FET device to control superconductivity and to modulate
the crystal structure, and open a new way to search for novel
electronic or structural phases. Because of its significant
advantage, the SIC-FET can serve in the modern electronic technology
to some degree.

This work is supported by the National Natural Science Foundation of
China (Grants No. 11190021, No. 11227902, No. 11534010 and No.
91422303), the Strategic Priority Research Program (B) of the
Chinese Academy of Sciences (Grant No. XDB04040100), the National Key R\&D
Program of the MOST of China (Grant No. 2016YFA0300201), and the Hefei
Science Center CAS (2016HSC-IU001).


\begin{thebibliography}{99}

\bibitem{18}J. G. Bednoz and K. A. M\"{u}ller. Z. Phys. B \textbf{64}, 189-193 (1986).
\bibitem{19}Y. Kamihara \emph{et al.}, J. Am. Chem. Soc. \textbf{130}, 3296-3297 (2008).
\bibitem{20}X. H. Chen \emph{et al.}, Nature \textbf{453},761-762 (2008).
\bibitem{1}R. E. Glover \emph{et al.}, Phys. Rev. Lett. \textbf{5}, 248-250 (1960).
\bibitem{2}C. H. Ahn \emph{et al.}, Science \textbf{284}, 1152-1155 (1999).
\bibitem{3}D. Chiba \emph{et al.}, Science \textbf{301}, 943 (2003).
\bibitem{4}C. H. Ahn \emph{et al.}, Nature \textbf{424}, 1015-1018 (2003).
\bibitem{5}K. S. Novoselov \emph{et al.}, Science \textbf{306}, 666 (2004).
\bibitem{6}A. D. Caviglia \emph{et al.}, Nature \textbf{456}, 624-627 (2008).
\bibitem{7}K. Ueno \emph{et al.}, Nat. Mater. \textbf{7}, 855 (2008).
\bibitem{8}K. Ueno \emph{et al.}, Nat. Nanotechnol. \textbf{6}, 408 (2011).
\bibitem{9}J. T. Ye \emph{et al.}, Nat. Mater. \textbf{9}, 125-128 (2010).
\bibitem{10}A. T. Bollinger \emph{et al.}, Nature \textbf{472}, 458-460 (2011).
\bibitem{11}J. T. Ye \emph{et al.}, Science \textbf{338}, 1193-1196 (2012).
\bibitem{12}Y. J. Yu \emph{et al.}, Nat. Nanotechnol. \textbf{10}, 270 (2015).
\bibitem{13}Y. Saito \emph{et al.}, Science \textbf{350}, 409-413 (2015).
\bibitem{14}J. M. Lu \emph{et al.}, Science \textbf{350}, 1353-1357 (2015).
\bibitem{15}L. J. Li \emph{et al.}, Nature \textbf{529}, 185-190 (2016).
\bibitem{21}B. Radisavljevic \emph{et al.}, Nat. Nanotechnol. \textbf{6}, 147-150 (2011).
\bibitem{22}L. K. Li \emph{et al.}, Nat. Nanotechnol. \textbf{9}, 372 (2014).
\bibitem{16}C. H. Ahn \emph{et al.}, Rev. Mod. Phys. \textbf{78}, 1185-1212 (2006).
\bibitem{17}K. Ueno \emph{et al.}, Journal of the Physical Society of Japan \textbf{83}, 032001 (2014).
\bibitem{23}H. T. Yuan \emph{et al.}, Adv. Funct. Mater. \textbf{19}, 1046-1053 (2009).
\bibitem{24}J. Jeong \emph{et al.}, Science \textbf{339}, 1402-1405 (2013).
\bibitem{25}T, D. Schladt \emph{et al.}, ACS Nano \textbf{7}, 8074-8081 (2013).
\bibitem{26}F. C. Hsu \emph{et al.}, Proc. Natl. Acad. Sci. USA \textbf{105}, 14262-14264 (2008).
\bibitem{27}J. G. Guo \emph{et al.}, Phys. Rev. B \textbf{82}, 180520(R) (2010).
\bibitem{28}A. F. Wang \emph{et al.}, Phys.Rev.B \textbf{83}, 060512(R) (2011).
\bibitem{29}M. Burrard-Lucas \emph{et al.}, Nat. Mater. \textbf{12}, 15-19 (2013).
\bibitem{30}X. F. Lu, \emph{et al.}, Nat. Mater. \textbf{14}, 325-329 (2015).
\bibitem{31}Q. Y. Wang \emph{et al.}, Chin. Phys. Lett. \textbf{29}, 037402 (2012).
\bibitem{32}D. F. Liu \emph{et al.}, Nat. Commun. \textbf{3}, 931 (2012).
\bibitem{33}S. L. He \emph{et al.}, Nat. Mater. \textbf{12}, 605-610 (2013).
\bibitem{34}S. Y. Tan \emph{et al.}, Nat. Mater. \textbf{12}, 634-640 (2013).
\bibitem{35}R. Peng \emph{et al.}, Nat. Commun. \textbf{5}, 5044 (2014).
\bibitem{36}G-F. Ge \emph{et al.}, Nat. Mater. \textbf{14}, 285-289 (2015).
\bibitem{37}L. Zhao \emph{et al.}, Nat. Commun. \textbf{7}, 10608 (2016).
\bibitem{38}X. H. Niu \emph{et al.}, Phys. Rev. B \textbf{92}, 060504(R) (2015).
\bibitem{39}B. Lei \emph{et al.}, Phys. Rev. Lett. \textbf{116}, 077002 (2016) .
\bibitem{40}K. Hanzawa \emph{et al.}, Proc. Natl. Acad. Sci. \textbf{113}, 3986-3990 (2016).
\bibitem{41}J. Guo \emph{et al.}, Nat. Commun. \textbf{5}, 4756 (2014)
\bibitem{42}X. D. Qi \emph{et al.}, Journal of Alloys and Compounds \textbf{509}, 6350鈥?353 (2011).
\bibitem{43}B. Lei \emph{et al.}, Phys. Rev. B \textbf{93}, 060501(R) (2016).
\bibitem{44}I. M. Lifshitz, Sov. Phys. JETP \textbf{11}, 1130 (1960).



\end{thebibliography}
\end{document}